\documentclass[se, manuscript]{copernicus}

\usepackage{amssymb}
\usepackage{amsmath}
\usepackage{bm}
\usepackage{booktabs}
\usepackage{hyperref}

\begin{document}
\nolinenumbers
\title{Learning Stratigraphically Consistent Relative Geologic Time from 3D Seismic Data via Sinusoidal Mapping}

% \Author[affil]{given_name}{surname}

\Author[1,2]{Yimin}{Dou}
\Author[1,2][xinmwu@ustc.edu.cn]{Xinming}{Wu}
\Author[1,2]{Hui}{Gao}
\Author[3]{Zhengfa}{Bi}

\affil[1]{State Key Laboratory of Precision Geodesy, School of Earth and Space Sciences, University of Science and Technology of China, Hefei, Anhui, China}
\affil[2]{Mengcheng National Geophysical Observatory, University of Science and Technology of China, Mengcheng, 233500, China}
\affil[3]{Earth \& Environmental Sciences, Energy Geosciences Division, Lawrence Berkeley National Laboratory, Berkeley, CA, USA}

%% The [] brackets identify the author with the corresponding affiliation. 1, 2, 3, etc. should be inserted.

\correspondence{Xinming Wu (xinmwu@ustc.edu.cn)}

\runningtitle{Learning Stratigraphically Consistent RGT via Sinusoidal Mapping}

\runningauthor{Y. Dou et al.}

\received{}
\pubdiscuss{} %% only important for two-stage journals
\revised{}
\accepted{}
\published{}

%% These dates will be inserted by Copernicus Publications during the typesetting process.

\firstpage{1}

\maketitle

\begin{abstract}
Relative Geologic Time (RGT) estimation from seismic data underpins subsurface structural modeling, depositional analysis, and reservoir characterization, providing the basis for horizon correlation and depositional system reconstruction. Accurate RGT estimation remains challenging because RGT is a topologically constrained continuous field in which local errors readily propagate globally through topological coupling, distorting the overall result. Conventional methods depend heavily on prior information, attribute extraction, and manual interaction, resulting in cumbersome workflows with limited automation. Existing deep-learning approaches predominantly adopt a regression formulation optimized by pixel-wise MSE/MAE losses, which struggles to recover thin horizons and fails to capture the stratigraphic semantics embedded in the RGT field, yielding limited generalization, unstable stratigraphic ordering, and poor adaptability to diverse structural and depositional settings.
We propose RGT-Est, a deep-learning framework that transfers the optimization target from the topologically constrained continuous field into a differentiable sinusoidal space. This representation explicitly encodes the periodic stratigraphic semantics of RGT and alleviates the over-smoothing of fine horizons inherent in direct regression. Pointwise, perceptual, and adversarial losses are jointly imposed in this space to enforce local fidelity, inter-layer consistency, and global structural plausibility, providing both fine-horizon discrimination and global stratigraphic awareness. An optional horizon-guidance module accepts sparse 2D or 3D horizons as priors to satisfy varying precision demands.
Trained on synthetic data and evaluated on field seismic surveys featuring dense faulting, large unconformities, steeply dipping strata, folded deformations, and clinoforms, RGT-Est achieves state-of-the-art performance among AI-based methods, and attains substantially higher horizon-correlation accuracy and topological consistency when sparse priors are incorporated. The code and data are available at \url{https://github.com/douyimin/RGT-Est} 
and \url{https://doi.org/10.5281/zenodo.20118902}.
\end{abstract}

\copyrightstatement{\copyright{} Author(s) 2026. This work is distributed under the Creative Commons Attribution 4.0 License.}

\introduction
\label{sec:intro}

Estimating the Relative Geologic Time (RGT) from seismic data is a fundamental task in seismic interpretation, carrying significant scientific and engineering value for subsurface structural modeling, depositional evolution analysis, reservoir characterization, and horizon correlation~\citep{zeng1998stratal,stark2003unwrapping,stark2004relative}. In an RGT volume, each sample is assigned a relative geologic time value.
Under normal stratigraphic conditions without overturned strata or reverse-fault-related structural repetition, RGT generally increases with depth; isosurfaces extracted from the RGT volume correspond to seismic horizons, while faults and unconformities are implicitly represented as discontinuities within the field~\citep{zeng1998stratal,stark2003unwrapping}. An accurate RGT volume can therefore encode, in a single representation, the horizons, faults, and unconformities present throughout the seismic volume, and has been successfully applied to downstream tasks such as sedimentologic interpretation, geobody detection, implicit structural modeling, and missing well-log prediction~\citep{zeng2012guidelines,grose2017structural,bader2018missing,karimi2015stratigraphic,geng2020deep,yang2023multi}.

\subsection{Conventional RGT Estimation Methods}
\label{sec:intro_conventional}

Conventional approaches to RGT estimation fall broadly into three categories. The first relies on manual horizon picking followed by interpolation: interpreters track as many horizons as possible by hand and interpolate them to construct an RGT volume~\citep{zeng1998stratal,debruin2007stratigraphic}. Although such methods honor the picked horizons exactly, the accuracy of the interpolated regions between horizons is difficult to guarantee, and the manual cost is prohibitive. The second category estimates RGT through seismic instantaneous phase unwrapping~\citep{stark2003unwrapping,wu2012generating}, exploiting the temporal information embedded in the instantaneous phase, but is prone to phase jumps and global distortion when the subsurface structure is complex or the signal-to-noise ratio is low. The third category is based on local dip or slope estimation~\citep{fomel2002applications,fomel2010predictive,lomask2006flattening,wu2017directional,wu2018leastsquares}, in which local reflector orientations are obtained via plane-wave destruction or structure tensors, and a global RGT field is subsequently reconstructed by least-squares fitting. While effective in regions of continuous reflections, these methods rely exclusively on local features and often fail to preserve horizon consistency across faults and unconformities~\citep{figueiredo2007seismic,lou2021simulating}. To alleviate this issue, several studies propose first ``unfaulting'' the seismic image before estimating the RGT~\citep{luo2013unfaulting,wu2016moving}, or inserting manually picked control points on both sides of faults as constraints~\citep{wu2015horizon}. However, all such strategies demand substantial prior information and manual interaction, resulting in cumbersome workflows with limited automation.

\subsection{Deep-Learning-Based RGT Estimation}
\label{sec:intro_dl}

In recent years, deep learning---in particular convolutional neural networks (CNNs)---has made substantial progress in seismic data processing~\citep{yu2018deep}, seismic interpretation~\citep{wu2019faultseg3d,gao2021channelseg3d,shi2019saltseg}, and seismic inversion~\citep{yang2019deep,geng2021deep}, opening new avenues for RGT estimation. \citet{geng2020deep} were the first to cast RGT estimation as a regression problem, proposing an end-to-end 2D CNN (a ResNet-50 encoder coupled with a U-Net-style decoder) that predicts RGT directly from a seismic section, optimized with the mean squared error (MSE) loss. Trained solely on synthetic data, their model generalizes well to field data containing complex faulting and folding, substantially outperforming predictive painting~\citep{wu2018leastsquares} and other local-filtering baselines. Building on this direction, \citet{bi2021deep} further coupled RGT estimation with fault detection using two separate 3D CNNs to address the two tasks in tandem, while \citet{yang2023multi} proposed a Transformer-based Multi-Task Learning (MTL) framework that unifies RGT estimation and fault detection in a single network: the two branches share feature maps and parameters, mutually constrain each other, and additionally accept interpreter-provided horizons as priors to improve robustness under complex fault systems, unconformities, and low-SNR data.

A separate but complementary line of work targets sparse-horizon picking and tracking, in which only a small number of stratigraphic surfaces are extracted from seismic data, typically driven by a few interpreter-provided seed points or sparse manual picks. To reduce the reliance on dense annotations, \citet{wu2022variable} proposed a weak-supervised framework that learns variable seismic waveform representations from sparse picks, and \citet{wang2022seismic} introduced a semi-supervised scheme with virtual adversarial training that propagates a few seed points to dense horizons and iteratively suggests regions lacking control points. To better handle structural discontinuities, \citet{li2022automatic} designed a network specifically tailored for tracking horizons across faults, while \citet{liu2024seismic} fused multiple seismic attributes through an ensemble dense inception transformer to improve robustness in low-SNR or weak-reflection zones. More recently, \citet{zhao2025seismic} adopted a TransUnet backbone that combines convolutional locality with Transformer-based long-range context for higher-fidelity horizon extraction, and \citet{liao2024deep} explicitly encoded interpretation uncertainty and a vertical ordering constraint into the training objective, noting that conventional point-by-point losses are ill-suited to horizon tracking and that domain shift further limits generalization to field data. These sparse-horizon methods can yield sharp, interpreter-controlled surfaces in regions where seed points are available, but they operate on one or a few horizons at a time and do not directly produce a globally consistent stratigraphic ordering across the entire volume. RGT estimation and sparse-horizon picking are therefore complementary: the former aims at a dense, volumetric stratigraphic representation, while the latter offers high-precision local control that can naturally serve as priors for the former---an interaction we exploit through the optional horizon-guidance module of RGT-Est.

Despite these advances, existing deep-learning approaches still face several fundamental bottlenecks, which we summarize below.

\subsection{Fundamental Limitations of Existing AI Methods}
\label{sec:intro_bottleneck}

\textit{Mismatch between the optimization target and the physical semantics of the RGT field.}
Existing AI methods predominantly formulate RGT estimation as continuous-field regression with MSE or MAE as the training objective~\citep{geng2020deep,bi2021deep}. Such pixel-wise losses implicitly assume that prediction errors at neighboring pixels are independent and identically distributed, effectively degrading the RGT field into a set of unrelated scalar predictions. This formulation overlooks the periodic and topological structure that RGT encodes as a ``stratigraphic code''---each isoline corresponds to a horizon and the layers follow a strict geologic time order. Under such losses, networks tend to converge to overly smoothed solutions that minimize the aggregate numerical error but fail to capture subtle inter-layer variations, resulting in blurred layer boundaries and suppressed thin-horizon signals, and thus a poor ability to resolve fine stratigraphic details in complex settings~\citep{yang2023multi}. In short, MSE/MAE optimization pursues ``numerical closeness'', whereas RGT estimation fundamentally requires ``stratigraphic correctness''---this mismatch is one of the key reasons why current AI methods struggle to push the accuracy ceiling further.

\textit{Inability to guarantee topological consistency.}
RGT is intrinsically a topologically constrained continuous field: it monotonically increases along depth, its isosurfaces never intersect, and it exhibits geologically specific patterns of discontinuity at faults and unconformities. This global topology implies that small local errors are never isolated: they propagate outward through monotonicity and stratigraphic coupling, eventually breaking the global ordering and producing horizon crossings, jumps, dislocations, or even inversions that are geologically implausible. Pure regression frameworks, however, lack any explicit modeling of such global topology---they can neither penalize topological violations at the loss level nor enforce stratigraphic constraints at the architectural level, but rely entirely on model capacity and data scale to ``implicitly learn'' the ordering rules. As a result, even small pointwise errors can translate into large-scale departures from the true stratigraphic order, and this implicit learning mechanism is particularly unstable in densely faulted zones, unconformities, and steeply dipping regions~\citep{geng2020deep}.

\textit{Limited generalization and structural adaptability.}
The two issues above jointly render current AI methods practically unusable for direct RGT estimation from seismic data in production. On the one hand, because the network merely ``memorizes'' stratigraphic patterns implicitly rather than understanding stratigraphic semantics at the representation level, its generalization depends heavily on the structural coverage of the training data: although performance on synthetic data within the training distribution is often satisfactory, it degrades sharply on unseen structural styles~\citep{geng2020deep}. On the other hand, the pixel-wise i.i.d.\ optimization assumption makes the network highly sensitive to input perturbations---noise, amplitude variations, frequency differences---and thus hampers transfer across surveys acquired with different parameters. Consequently, existing AI methods for RGT are mostly demonstrated on carefully curated synthetic datasets or on small, simply structured field slices~\citep{geng2020deep,yang2023multi,bi2021deep}, while convincing large-scale validation on full 3D field volumes, across different surveys, and in structurally complex regions remains unreported.

\subsection{Our Work}
\label{sec:intro_contrib}

To address the above issues, we propose RGT-Est, a deep learning framework that learns relative geologic time from 3D seismic data through a sinusoidal mapping. In contrast to existing regression-based methods, we no longer treat the RGT field as a scalar field to be pixel-wise fitted; instead, the scalar field output by the network is mapped into a differentiable sinusoidal space for modeling and optimization. The sinusoidal space inherently encodes the periodic stratigraphic semantics of RGT: we adopt three sinusoidal channels with linearly decreasing frequencies to build a multi-scale phase encoding of the RGT---the high-frequency channel captures fine-scale horizons such as thin layers, the low-frequency channel encodes large-scale stratigraphic framework, and the combination of the three channels yields a unique and distinguishable representation of any RGT value, thereby simultaneously accommodating local detail and global stratigraphic structure within a unified differentiable framework. In this sinusoidal space, subtle stratigraphic differences are amplified into clearly distinguishable phase variations, allowing loss functions to operate in a more discriminative feature space. Based on this representation, we jointly impose adversarial, perceptual, and MAE losses in the sinusoidal space: the adversarial loss drives the prediction to approach the real stratigraphic distribution globally, the perceptual loss enforces structural consistency across multi-scale features, and the MAE loss stabilizes pointwise numerical fitting. Together, they balance global topology, structural semantics, and local accuracy, so that the approach captures thin-layer details while retaining a robust global stratigraphic view.

The main contributions of this work are summarized as follows.
\begin{enumerate}
	\item \textbf{A sinusoidal-space modeling paradigm for RGT estimation.}
	We reformulate RGT estimation from continuous scalar-field regression into a multi-scale phase optimization problem in a sinusoidal space. Three sinusoidal channels with linearly decreasing frequencies explicitly encode the periodic stratigraphic semantics of RGT and yield a unique representation of any RGT value, fundamentally alleviating the over-smoothing of thin layers caused by MSE/MAE losses.
	
	\item \textbf{A multi-loss collaborative mechanism for global topological constraints.}
	We jointly impose adversarial, perceptual, and MAE losses in the sinusoidal space, constraining the network from three complementary perspectives---distributional consistency, structural fidelity, and pointwise accuracy---and equipping it with both fine-horizon discrimination and robust global stratigraphic awareness.
	
	\item \textbf{Optional sparse horizon guidance.}
	We design an optional Horizon Guidance module that accepts sparse 2D or 3D horizons as priors. RGT-Est operates fully automatically without any prior; once horizons are provided, it delivers substantially higher precision and naturally preserves lateral consistency in slice-by-slice 3D prediction.
	
	\item \textbf{Systematic multi-scenario generalization evaluation.}
	We evaluate RGT-Est on multiple structurally complex field seismic datasets covering unconformities, densely faulted systems, steeply dipping structures, and strong structural superposition. RGT-Est substantially outperforms publicly available AI-based RGT estimation method---producing high-quality RGT fields end-to-end in direct inference mode, and reaching very high precision once sparse horizons are provided---validating its strong generalization across surveys, structural styles, and data qualities.
\end{enumerate}

\section{Method}
\label{sec:method}

\subsection{Overview}
\label{sec:method_overview}

\begin{figure*}[t]
\includegraphics[width=17cm]{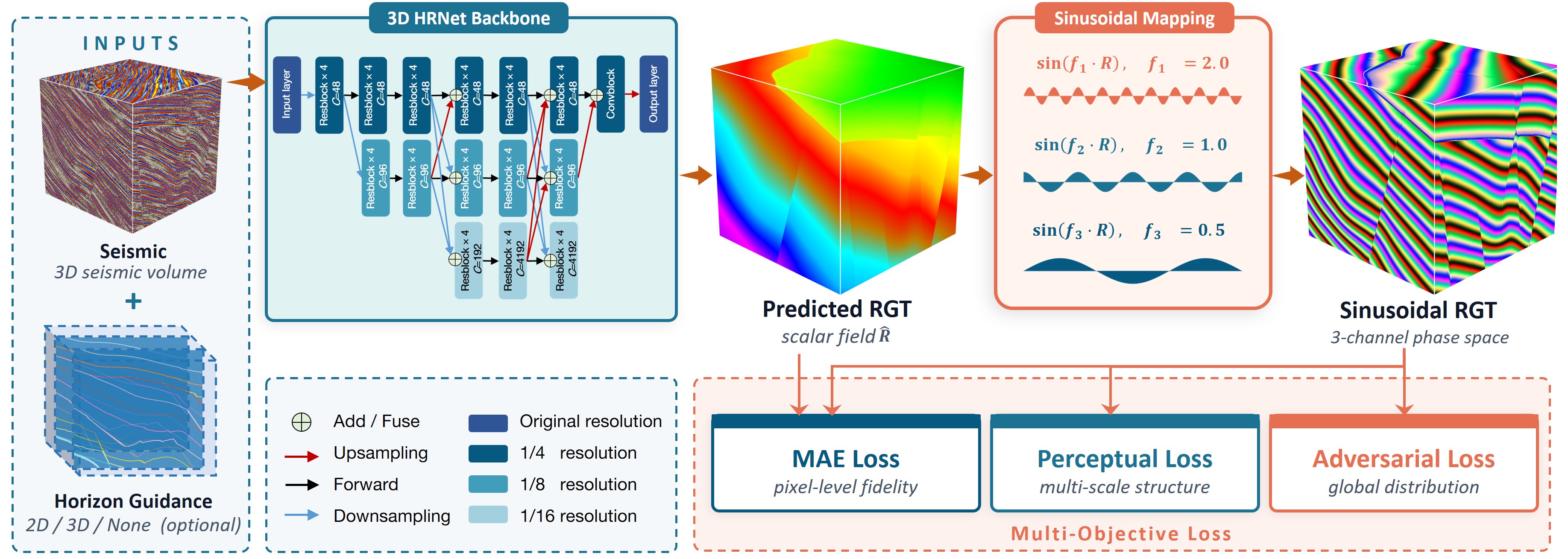}
\caption{Overview of the proposed RGT-Est framework. A 3D HRNet backbone takes a 3D seismic volume with optional 2D / 3D horizon guidance as input and predicts a continuous RGT scalar field $\hat{R}$. The Sinusoidal Mapping module maps $\hat{R}$ into a three-channel phase space via $\sin(f_i \hat{R})$ with $f_1{=}2.0$, $f_2{=}1.0$, $f_3{=}0.5$, yielding the Sinusoidal RGT. Training jointly minimizes the MAE, perceptual, and adversarial losses over the Predicted RGT and the Sinusoidal RGT representations.}
\label{fig:framework}
\end{figure*}

The overall framework of RGT-Est is illustrated in Fig.~\ref{fig:framework}. The input consists of a 3D seismic volume $\bm{S} \in \mathbb{R}^{1 \times D \times H \times W}$ together with an optional sparse horizon guidance $\bm{H} \in \mathbb{R}^{1 \times D \times H \times W}$ (2D, 3D, or empty), accompanied by an indicator mask $\bm{M} = \mathbb{I}(\bm{H} \neq 0)$ that marks the positions of user-provided horizons. When no prior is available, $\bm{H}$ and $\bm{M}$ are simply set to zero, so that the method naturally supports both fully automatic inference and sparse-horizon-guided modes within a single architecture. A 3D HRNet serves as the backbone $G_\theta$ and predicts, in an end-to-end manner, a continuous RGT scalar field $\hat{R} \in [-1, 1]^{1 \times D \times H \times W}$ through
\begin{equation}
	\hat{R} = \tanh \bigl( G_\theta(\bm{S}, \bm{H}, \bm{M}) \bigr).
	\label{eq:forward}
\end{equation}
The prediction $\hat{R}$ is then fed into the \emph{Sinusoidal Mapping} module $\mathcal{T}$, which maps the scalar field into a three-channel phase space via frequency-decreasing sinusoidal encodings; the resulting Sinusoidal RGT amplifies subtle stratigraphic differences and explicitly expresses the periodic stratigraphic semantics embedded in $\hat{R}$. A 3D PatchGAN discriminator $D_\phi$ is introduced to distinguish the predicted phase representation from the ground-truth one, conditioned on the input seismic volume. During training, three complementary losses---a hybrid regression loss, a multi-directional perceptual loss, and a relativistic adversarial loss---are jointly minimized to reconcile local accuracy, structural semantics, and global topology.

\subsection{Sinusoidal Mapping}
\label{sec:method_sinusoidal}

A key innovation of RGT-Est is that losses are \emph{not} evaluated solely on the scalar RGT field $\hat{R}$; instead, $\hat{R}$ is first lifted into a higher-dimensional phase space through a differentiable Sinusoidal Mapping module $\mathcal{T}$. Given $\hat{R} \in [-1, 1]$, we first rescale it to the discretized range $[1, N]$ with $N = 256$:
\begin{equation}
	R = \frac{\hat{R}+1}{2}\,(N-1) + 1,
	\label{eq:rescale}
\end{equation}
and then construct a three-channel phase representation using sinusoidal basis functions with linearly decreasing frequencies:
\begin{equation}
	\mathcal{T}(R) \;=\; \bigl[\, \sin(f_1 R),\; \sin(f_2 R),\; \sin(f_3 R) \,\bigr]^{\top},
	\qquad (f_1, f_2, f_3) = (2.0,\; 1.0,\; 0.5).
	\label{eq:sinusoidal}
\end{equation}
The geometric frequency decay $(f_1 : f_2 : f_3 = 4 : 2 : 1)$ is chosen so that the three channels jointly span a wide range of stratigraphic scales: the high-frequency channel $\sin(f_1 R)$ captures thin layers and fine-scale horizons, the low-frequency channel $\sin(f_3 R)$ encodes the large-scale stratigraphic framework, and the mid-frequency channel interpolates between the two. In this multi-frequency space every RGT value is mapped to a unique, distinguishable three-dimensional phase vector, preventing the wrap-around ambiguity that would arise from a single-frequency encoding. Moreover, minute variations in $R$ are amplified into clearly separable phase differences in $\mathcal{T}(R)$, allowing downstream losses to operate in a more discriminative feature space. Throughout the remainder of the paper, we denote the predicted and ground-truth phase representations by
\begin{equation}
	\hat{\bm{P}} \;=\; \mathcal{T}(\hat{R}),
	\qquad
	\bm{P} \;=\; \mathcal{T}(R).
	\label{eq:phase_notation}
\end{equation}

\begin{figure*}[t]
\includegraphics[width=17cm]{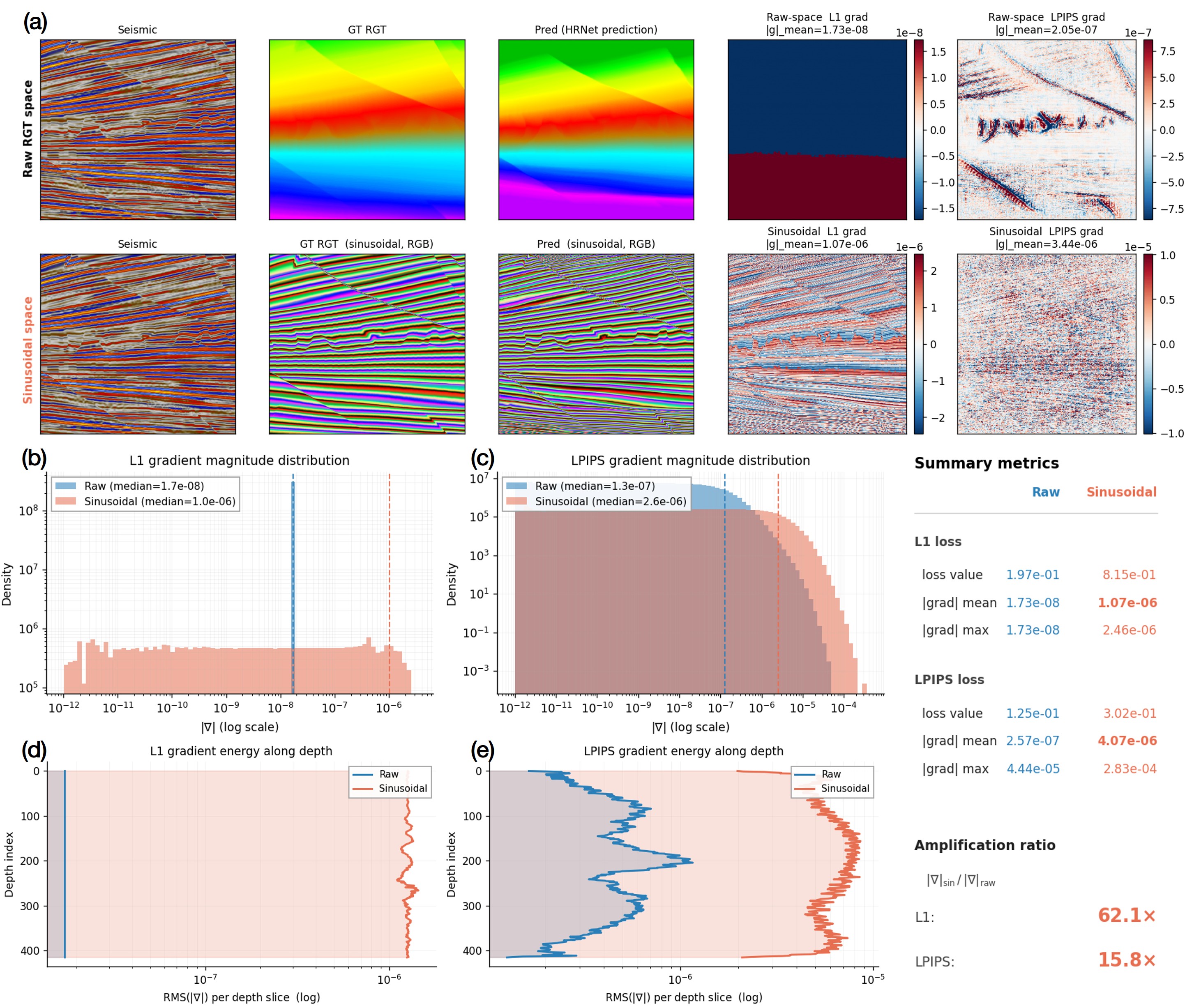}
\caption{Gradient back-propagation comparison between the raw RGT space and the proposed sinusoidal space. A pretrained 3D~HRNet is kept frozen, and the $L_1$ and 3D~LPIPS losses are computed either directly on $\hat{R}$ or on its sinusoidal encoding $\mathcal{T}(\hat{R})$; the gradient at $\hat{R}$ is recorded for comparison. (a)~Inline slice of the seismic input, the GT RGT, the prediction, and the resulting $L_1$/LPIPS gradient maps in both domains. (b,c)~Log-scale per-voxel distributions of $|\nabla|$. (d,e)~Depth-wise RMS of $|\nabla|$. Right: summary statistics and the overall amplification ratio $|\nabla|_{\mathrm{sin}}/|\nabla|_{\mathrm{raw}}$ ($62.1\times$ for $L_1$, $15.8\times$ for LPIPS).}
\label{fig:grad_analysis}
\end{figure*}

To empirically verify the mechanism behind the proposed Sinusoidal Mapping, we perform a gradient back-propagation analysis that compares the training signal received by the prediction under the raw scalar representation versus the sinusoidal phase representation (Fig.~\ref{fig:grad_analysis}). We freeze a pretrained 3D~HRNet, feed a real seismic volume, and take its forward output $\hat{R}$ as the prediction; the ground-truth RGT $R$ is retained as the regression target. Two losses---the $L_1$ loss and the multi-directional 3D~LPIPS loss---are computed in two equivalent domains: (i)~directly on the scalar RGT field, and (ii)~on the three-channel phase field $\mathcal{T}(\cdot) = [\sin(f_1 R), \sin(f_2 R), \sin(f_3 R)]$ with $(f_1, f_2, f_3) = (2.0, 1.0, 0.5)$ as defined in Sect.~\ref{sec:method_sinusoidal}. For every configuration the backward pass is traced to the prediction, and the gradient tensor at $\hat{R}$ is recorded and compared.

Figure~\ref{fig:grad_analysis}a contrasts the two domains on a central inline slice. The raw-space $L_1$ gradient collapses to a nearly piecewise-constant sign map ($|g|_{\text{mean}}{=}1.73\!\times\!10^{-8}$), because the pixelwise $\mathcal{L}_{L_1} = |\hat{R} - R|$ produces a gradient of magnitude $1/N$ that is completely blind to whether the error occurs at a thin horizon or a homogeneous interior. More revealingly, in panel~(a) this raw-space $L_1$ gradient exhibits a sharp positive--negative interface that partitions the slice into two nearly uniform regions, indicating that the regression signal is merely pulling $\hat{R}$ toward or away from the bulk depth trend rather than encoding any stratigraphic structure; the raw-space LPIPS gradient suffers from the same limitation to a milder extent, responding only to a few high-amplitude reflectors and leaving the vast majority of layer interiors essentially unsupervised. In sharp contrast, the sinusoidal-space gradients of both $L_1$ and LPIPS exhibit strong, spatially coherent responses that trace the underlying stratigraphic layering---exactly the fine-scale horizons that conventional regression losses tend to smooth out. The gradient magnitude increases by orders of magnitude, confirming that the same prediction error produces a substantially stronger and more discriminative training signal once it is measured in the sinusoidal phase space.

Figure~\ref{fig:grad_analysis}b,c plot the per-voxel distributions of $|\nabla|$ on a logarithmic scale. The raw-space distributions are narrowly concentrated near $10^{-8}$--$10^{-7}$ with a heavy truncation at the lower tail (the gradient magnitude is bounded by $1/N$ for $L_1$). The sinusoidal-space distributions are shifted by roughly one-to-two decades to the right and exhibit a much longer tail, with medians moving from $1.7\!\times\!10^{-8}$ to $1.0\!\times\!10^{-6}$ for $L_1$ and from $1.3\!\times\!10^{-7}$ to $2.6\!\times\!10^{-6}$ for LPIPS. A richer tail indicates that more voxels receive informative, non-saturating gradient updates, which translates into finer-grained parameter updates during training.

Figure~\ref{fig:grad_analysis}d,e show the root-mean-square of $|\nabla|$ computed per depth slice. For raw-space $L_1$, the profile is an almost flat vertical line because the loss degenerates to sign-based updates and has no depth-dependent structure. For the sinusoidal-space counterpart, the profile exhibits clear stratigraphic modulation---larger gradients at layer boundaries and smaller gradients in between---which is precisely the behavior required to correct thin-horizon errors without over-smoothing the interior. The LPIPS profiles show the same trend but at a higher baseline.

Across the volume, computing the loss in the sinusoidal space amplifies the mean gradient magnitude received by the prediction by $62.1\times$ for $L_1$ and $15.8\times$ for LPIPS (right panel of Fig.~\ref{fig:grad_analysis}). This is consistent with the first-order estimate
\begin{equation}
	\frac{\partial \mathcal{T}_c(R)}{\partial R}
	\;=\; f_c \cdot \tfrac{N-1}{2} \cdot \cos(f_c\,s),
	\label{eq:grad_amp}
\end{equation}
where $s \in [1, N]$ is the rescaled RGT value and $N{=}256$: the product $f_c \cdot (N-1)/2$ already yields an amplification factor on the order of $10^2$ per channel, which is then modulated by the local phase term $\cos(f_c\,s)$ and redistributes the energy toward layer boundaries. The sinusoidal representation therefore acts as a \emph{differentiable, stratigraphy-aware amplifier} of the training signal, explaining why the multi-objective losses defined in Sect.~\ref{sec:method_loss} are particularly effective at recovering thin horizons and preserving stratigraphic ordering.

\subsection{3D HRNet Backbone}
\label{sec:method_backbone}

We adopt a three-dimensional High-Resolution Network (HRNet) as the backbone $G_\theta$ because its parallel multi-resolution branches are well suited for RGT estimation: the high-resolution branch preserves fine-scale horizon details, while the low-resolution branches enlarge the receptive field so that global stratigraphic structure can be captured. Given the concatenated input $[\bm{S}, \bm{H}, \bm{M}]$ of three channels, a two-layer strided convolution stem first downsamples the volume by a factor of four. Stage~1 is a sequence of four Bottleneck residual blocks operating at $1/4$ resolution with 256 channels; Transition~1 then branches out an additional $1/8$-resolution stream with $2c$ channels ($c=48$ in all experiments). Stage~2 stacks two StageModules, each containing four BasicBlock residual units per branch and a cross-resolution fusion that exchanges information between the two parallel streams. Transition~2 further introduces a $1/16$-resolution branch with $4c$ channels, and Stage~3 contains six StageModules that repeatedly fuse the three parallel streams through learnable upsampling, downsampling, and identity connections. After Stage~3 the highest two resolutions ($1/4$ and $1/8$) are retained, concatenated after bilinear upsampling, and fed into a regression decoder consisting of $3 \times 3 \times 3$ convolutions interleaved with two $\mathrm{PixelShuffle3D}$ layers that progressively recover the original spatial resolution. A final $\tanh$ activation constrains the predicted RGT to $[-1, 1]$, matching the normalization range of the training labels and stabilizing the subsequent sinusoidal transform.

\subsection{Discriminator Architecture}
\label{sec:method_disc}

To impose a distribution-level topological constraint on the predicted RGT, we employ a 3D PatchGAN discriminator~\citep{isola2017image} $D_\phi$ conditioned on the input seismic volume. $D_\phi$ takes as input a four-channel volume formed by concatenating the phase representation (three channels) with the seismic volume (one channel), and outputs a dense patch-wise logit map. Concretely, $D_\phi$ consists of three strided $4 \times 4 \times 4$ convolutional blocks with base width $b=100$ and a stride of $2$, each (except the first) followed by 3D instance normalization and a LeakyReLU activation with negative slope $0.2$, and is terminated by a $4 \times 4 \times 4$ convolution with unit stride that produces the patch logits. The patch-wise design allows $D_\phi$ to judge local stratigraphic plausibility across the entire volume, rather than committing to a single global decision.

\subsection{Loss Functions}
\label{sec:method_loss}

RGT-Est is trained by minimizing a weighted combination of three complementary losses operating on both the scalar field $\hat{R}$ and the phase representation $\hat{\bm{P}}$.

\subsubsection*{Hybrid regression loss}
A hybrid $L_1$/$L_2$ loss $\ell$ is first applied on the scalar RGT field to enforce pointwise numerical fidelity:
\begin{equation}
	\ell(a, b) \;=\; \lVert a - b \rVert_1 \;+\; 0.5\,\lVert a - b \rVert_2^2.
	\label{eq:logreg}
\end{equation}
Building on this, a regression loss that couples the scalar and phase domains is defined as
\begin{equation}
	\mathcal{L}_{\text{reg}} \;=\; 2\,\ell\!\bigl(\hat{R},\, R\bigr) \;+\; 0.2\,\lVert \hat{\bm{P}} - \bm{P} \rVert_1.
	\label{eq:reg_loss}
\end{equation}
The scalar term stabilizes large-scale fitting while the phase term amplifies the penalty on fine-scale stratigraphic errors, which are hard to recover from pure scalar regression.

\subsubsection*{Multi-directional 3D perceptual loss}
To enforce multi-scale structural consistency, we extend the 2D LPIPS~\citep{zhang2018unreasonable} metric to three dimensions by means of a \emph{multi-directional slicing} strategy. For any two volumes $\bm{X}, \bm{Y} \in \mathbb{R}^{N \times C \times D \times H \times W}$, we uniformly sample $k=32$ slice indices along each axis and compute the 2D LPIPS similarity on the extracted slices, which are then averaged across all three directions:
\begin{equation}
	\mathcal{L}_{\text{per}}^{3\mathrm{D}}(\bm{X}, \bm{Y})
	\;=\; \frac{1}{3} \sum_{d \in \{D, H, W\}} \frac{1}{k} \sum_{i \in \mathcal{I}_d}
	\mathrm{LPIPS}_{\text{AlexNet}}\!\bigl(\bm{X}^{(d)}_i, \bm{Y}^{(d)}_i\bigr),
	\label{eq:lpips3d}
\end{equation}
where $\mathcal{I}_d$ denotes the sampled indices along axis $d$ and $\bm{X}^{(d)}_i$ denotes the $i$-th 2D slice of $\bm{X}$ along axis $d$. The overall perceptual loss is the sum of the perceptual distances computed on the phase representation and on the scalar field:
\begin{equation}
	\mathcal{L}_{\text{lpips}}
	\;=\; \mathcal{L}_{\text{per}}^{3\mathrm{D}}\!\bigl(\hat{\bm{P}},\, \bm{P}\bigr)
	\;+\; \mathcal{L}_{\text{per}}^{3\mathrm{D}}\!\bigl(\hat{R},\, R\bigr).
	\label{eq:lpips_total}
\end{equation}
This formulation encourages the network to match the ground truth not only pointwise but also in terms of high-level perceptual structure along all three spatial axes.

\subsubsection*{Relativistic hinge adversarial loss}
To impose a distribution-level topological constraint, we adopt the relativistic hinge adversarial formulation of \citet{jolicoeur2018relativistic}, which combines the hinge GAN loss~\citep{lim2017geometric} with the relativistic discriminator scheme. Let $\hat{\bm{P}}$ and $\bm{P}$ denote the predicted and ground-truth phase representations, both conditioned on the seismic input $\bm{S}$. Writing $d_r = D_\phi([\bm{P}, \bm{S}])$ and $d_f = D_\phi([\hat{\bm{P}}, \bm{S}])$, the discriminator loss is
\begin{equation}
	\begin{aligned}
		\mathcal{L}_D \;=\; &\tfrac{1}{2}\,\mathbb{E}\bigl[\,\mathrm{ReLU}\!\bigl(1 - (d_r - \mathbb{E}[d_f])\bigr)\,\bigr] 
		+\; \tfrac{1}{2}\,\mathbb{E}\bigl[\,\mathrm{ReLU}\!\bigl(1 + (d_f - \mathbb{E}[d_r])\bigr)\,\bigr],
	\end{aligned}
	\label{eq:disc_loss}
\end{equation}
and the generator adversarial loss is correspondingly
\begin{equation}
	\begin{aligned}
		\mathcal{L}_{\text{adv}} \;=\; &\tfrac{1}{2}\,\mathbb{E}\bigl[\,\mathrm{ReLU}\!\bigl(1 + (d_r - \mathbb{E}[d_f])\bigr)\,\bigr] 
		+\; \tfrac{1}{2}\,\mathbb{E}\bigl[\,\mathrm{ReLU}\!\bigl(1 - (d_f - \mathbb{E}[d_r])\bigr)\,\bigr].
	\end{aligned}
	\label{eq:gen_adv_loss}
\end{equation}
Relativistic losses compare each real/fake sample against the expected opposite, which is known to yield more informative gradients and more stable training than the non-saturating or hinge variants. Because $\mathcal{L}_{\text{adv}}$ is computed in the phase space, the discriminator effectively regularizes the \emph{stratigraphic distribution} of the prediction rather than its raw scalar values.

\subsubsection*{Total generator objective}
The generator is trained with the weighted sum
\begin{equation}
	\mathcal{L}_G \;=\; \lambda_{\text{reg}}\,\mathcal{L}_{\text{reg}}
	\;+\; \lambda_{\text{lpips}}\,\mathcal{L}_{\text{lpips}}
	\;+\; \lambda_{\text{adv}}\,\mathcal{L}_{\text{adv}},
	\label{eq:total_g}
\end{equation}
with $\lambda_{\text{reg}} = 5$, $\lambda_{\text{lpips}} = 5$, and $\lambda_{\text{adv}} = 0.01$. The discriminator is trained with $\lambda_{\text{adv}} \mathcal{L}_D$ using the same adversarial weight.

\subsection{Training Data}
\label{sec:training_data}

To train RGT-Est, we randomly generated 2,000 pairs of synthetic 3D volumes \citep{wu2020building}, each with a size of $512 \times 512 \times 512$ (an example is shown in Fig.~\ref{fig:framework}).
Each pair consists of a seismic amplitude volume and its corresponding ground-truth RGT volume.
The synthetic models were designed to cover a wide range of structural and stratigraphic scenarios that commonly challenge RGT estimation, including normal and reverse faults, densely faulted systems, unconformities, complex folds, slope structures, and laterally varying depositional architectures.
These geological elements introduce reflector discontinuities, abrupt stratigraphic terminations, strong lateral variations in dip, and multi-scale deformation patterns, thereby providing diverse training examples for learning stratigraphically consistent RGT representations. The corresponding seismic volumes were generated from the synthetic RGT models through forward modeling, so that the seismic reflections are consistent with the underlying stratigraphic ordering.

\subsection{Optimization and Implementation}
\label{sec:method_optim}

The generator $G_\theta$ and discriminator $D_\phi$ are optimized alternately in a two-player minimax game. In each iteration, we first update $G_\theta$ by minimizing $\mathcal{L}_G$ with $D_\phi$'s parameters frozen, and then update $D_\phi$ by minimizing $\lambda_{\text{adv}} \mathcal{L}_D$ on detached generator outputs. Both networks are optimized with the Adam optimizer~\citep{kingma2015adam} using $\beta_1 = 0$, $\beta_2 = 0.9$, and a constant learning rate of $10^{-4}$, a configuration empirically shown to stabilize relativistic-hinge GAN training. To enlarge data diversity at negligible cost, at each training step we randomly swap the last two spatial axes of the seismic, target RGT, and horizon-guidance volumes with probability $0.5$, which acts as an isotropic in-plane augmentation. All experiments are implemented in PyTorch~Lightning with mixed-precision training on NVIDIA A100 80G $\times$ 4 GPUs. Batch size is set to 4, and the training input dimension is $400 \times 512 \times 256$ (timeline, inline, crossline).

\section{Results}
\label{sec:results}

We conduct three sets of experiments to evaluate RGT-Est. \textbf{First}, we isolate the contribution of the sinusoidal mapping by directly comparing optimization in the proposed phase space with optimization in the raw RGT voxel space---the formulation used by DeepRGT~\citep{bi2021deep} and representative of existing AI-based RGT estimation methods. To ensure a fair comparison, the backbone and training data are held identical across the two settings. \textbf{Second}, we examine how sparse 2D (single-section) and 3D (volumetric) horizon priors progressively improve inference accuracy in structurally complex regions. \textbf{Third}, we test the generalization of RGT-Est on multiple highly challenging field surveys featuring densely faulted zones, large unconformities, steeply dipping strata, folded deformations, and clinoforms.

\subsection{Comparison with Voxel-Space Optimization}
\label{sec:exp_comparison}

To assess the contribution of the proposed sinusoidal mapping, we compare RGT-Est against a re-implemented version of DeepRGT~\citep{bi2021deep}, the most representative AI-based RGT estimation method that performs optimization directly in the raw RGT voxel space. To remove confounding factors arising from differences in network capacity or training data, we re-implement DeepRGT by replacing its original backbone with the same 3D HRNet used in this work and retraining it on the same training set as RGT-Est, while keeping the rest of its formulation---namely the voxel-space regression loss, optimization schedule, augmentation strategy, and inference protocol---unchanged. Hereafter we refer to this baseline as DeepRGT$^\dagger$ to indicate that it shares the backbone and training data with RGT-Est. We evaluate both methods on four field surveys with relatively mild structural deformation: the Costa Rica fore-arc basin and three publicly available Netherlands datasets. These surveys span a range of acquisition geometries and stratigraphic styles while remaining within a regime where pixelwise regression baselines are expected to behave reliably, thereby providing a fair benchmark for examining the additional value brought by sinusoidal-space optimization. For each survey, we present representative inline and time-slice visualizations that highlight the differences in fine-scale layering and global structural coherence between the two methods.

\begin{figure*}[t]
\includegraphics[width=17cm]{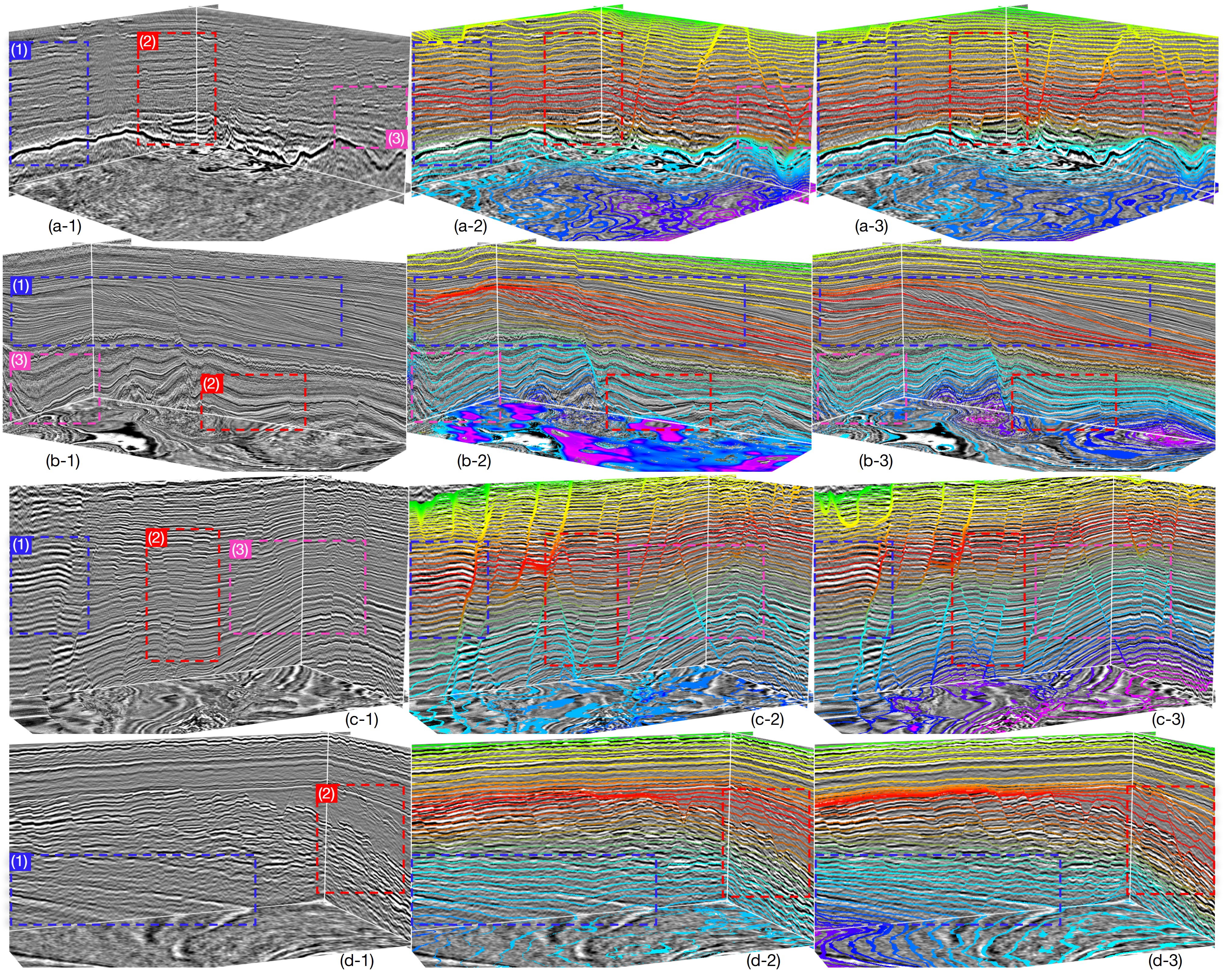}
\caption{Qualitative comparison of RGT estimation on field seismic volumes. From left to right, each group presents the input seismic volume, the result of DeepRGT$^\dagger$ (re-implementation) \citep{bi2021deep}, and the result of the proposed RGT-Est. For both methods, the displayed horizons are extracted as iso-surfaces from the estimated RGT fields and overlaid on the seismic volumes for visual comparison. The colored dashed boxes highlight challenging regions, including weak reflectors, slope structures, faulted zones, and laterally varying stratigraphy. RGT-Est better follows seismic reflectors and preserves more coherent stratigraphic ordering.}
\label{fig:comparison}
\end{figure*}

Figure~\ref{fig:comparison} presents the field-data comparison between DeepRGT$^\dagger$ and the proposed RGT-Est on the Costa Rica and Netherlands datasets.
The first row shows the Costa Rica data, whereas the remaining three rows show examples from the Netherlands data.
For each example, the first column shows the input seismic volume, the second column shows the result of DeepRGT$^\dagger$, and the third column shows the result of RGT-Est.
Because reliable RGT estimation requires both local horizon fidelity and global stratigraphic ordering, we mainly evaluate the results from four aspects: alignment between RGT contours and seismic reflectors, preservation of thin layers, continuity across laterally varying structures, and topological consistency in complex deformation zones.

In the Costa Rica example shown in the first row, the seismic image contains weak reflectors, discontinuous events, and strong structural deformation.
These features make RGT estimation difficult because local errors can easily propagate into large-scale distortions of the stratigraphic field.
DeepRGT$^\dagger$ recovers the general depth-increasing trend, but the estimated contours show clear instability in the highlighted regions.
In the blue box, where the seismic reflections are relatively weak and closely spaced, the contours produced by DeepRGT$^\dagger$ are over-smoothed and fail to sufficiently resolve fine stratigraphic variations.
In the red box, where the reflectors are strongly deformed, the contours become locally distorted and are less consistent with the seismic events.
The magenta box further shows that DeepRGT$^\dagger$ has difficulty maintaining stable layer continuity in regions with lateral structural changes.
By contrast, RGT-Est produces contours that are more consistent with the reflector geometry.
The predicted RGT field better follows the deformed seismic events, preserves more detailed stratigraphic variations, and maintains a more coherent ordering from shallow to deep intervals.
This result indicates that RGT-Est generalizes well to structurally complex field data with weak and discontinuous reflectors.

The second row shows the Netherlands F3 dataset, where a prominent slope structure is present in the shallow interval, as highlighted by the blue box.
This case is challenging because the reflectors within the slope are laterally inclined and locally thinned, requiring the estimated RGT contours to follow both the dipping geometry and the internal stratigraphic order.
DeepRGT$^\dagger$ captures the overall stratigraphic trend, but its contours in the slope zone are relatively over-smoothed and do not fully conform to the inclined seismic reflectors.
As a result, the internal layering of the slope is weakened, and the lateral variation of the depositional architecture is not sufficiently preserved.
In the deeper red-box region, where the reflectors become weaker and more structurally complicated, DeepRGT$^\dagger$ also produces locally unstable contours.
In comparison, RGT-Est better preserves the internal geometry of the slope.
Its contours remain more parallel to the inclined reflectors and show stronger lateral continuity across the shallow depositional structure.
In the deeper complex region, RGT-Est also gives a more coherent stratigraphic field, suggesting improved robustness to weak reflections and local structural disturbance.

The third row further evaluates the methods on a Netherlands example with stronger structural variation and multiple fault-related discontinuities.
In the input seismic volume, the blue, red, and magenta boxes mark regions with different interpretation difficulties, including weakly expressed strata, faulted or steeply varying reflectors, and laterally changing seismic patterns.
DeepRGT$^\dagger$ gives a reasonable large-scale RGT trend, but its contours become less reliable in these complex regions.
In particular, around the central faulted zone marked by the red box, the contours tend to be locally blurred and poorly aligned with the displaced reflectors.
In the magenta-box region, the estimated contours are also less consistent with the lateral changes in reflector geometry.
These artifacts suggest that direct regression tends to suppress subtle stratigraphic differences and may produce locally inconsistent RGT fields when the seismic structures depart from simple layered patterns.
RGT-Est substantially improves the result.
The predicted contours better track the seismic reflectors on both sides of the faulted zone and preserve clearer stratigraphic separation across laterally varying structures.
The result shows fewer local distortions and a more stable ordering of the RGT field, demonstrating the advantage of the proposed stratigraphy-aware optimization.

The fourth row shows another Netherlands example characterized by broad weak-reflection intervals and local deformation.
The blue box covers a large region where the reflectors are relatively subtle and laterally continuous, whereas the red box highlights a structurally complicated zone with rapid changes in reflector geometry.
DeepRGT$^\dagger$ tends to produce overly smooth contours in the weak-reflection interval, which reduces the resolution of fine stratigraphic layers.
Near the red-box region, its contours also show local distortion and insufficient conformity with the seismic events.
RGT-Est produces a more geologically plausible RGT field.
In the blue-box region, the contours are more continuous and better aligned with the subtle reflectors, indicating improved sensitivity to weak but coherent stratigraphic signals.
Around the red-box region, RGT-Est better adapts to the local deformation and preserves a more consistent relationship between adjacent layers.

Overall, the field-data comparisons demonstrate that RGT-Est consistently outperforms DeepRGT$^\dagger$ across different surveys and structural settings.
DeepRGT$^\dagger$ can recover the first-order stratigraphic trend, but it often suffers from over-smoothing, weakened thin-layer discrimination, and local contour distortion in weak-reflection zones, slope structures, faulted regions, and laterally varying depositional architectures.
In contrast, RGT-Est produces RGT contours that are more consistent with seismic reflectors and more stable across complex structural transitions.
The improvement is particularly evident in the highlighted regions, where the proposed method better preserves fine stratigraphic details while maintaining global topological consistency.
These observations support the effectiveness of transferring RGT optimization from the raw scalar space to the sinusoidal phase space, where subtle inter-layer differences can be more explicitly represented and jointly constrained by pointwise, perceptual, and adversarial losses.

\subsection{Stratigraphic-Constraint-Guided RGT Estimation}
\label{sec:exp_constraint}

\begin{figure*}[t]
\includegraphics[width=17cm]{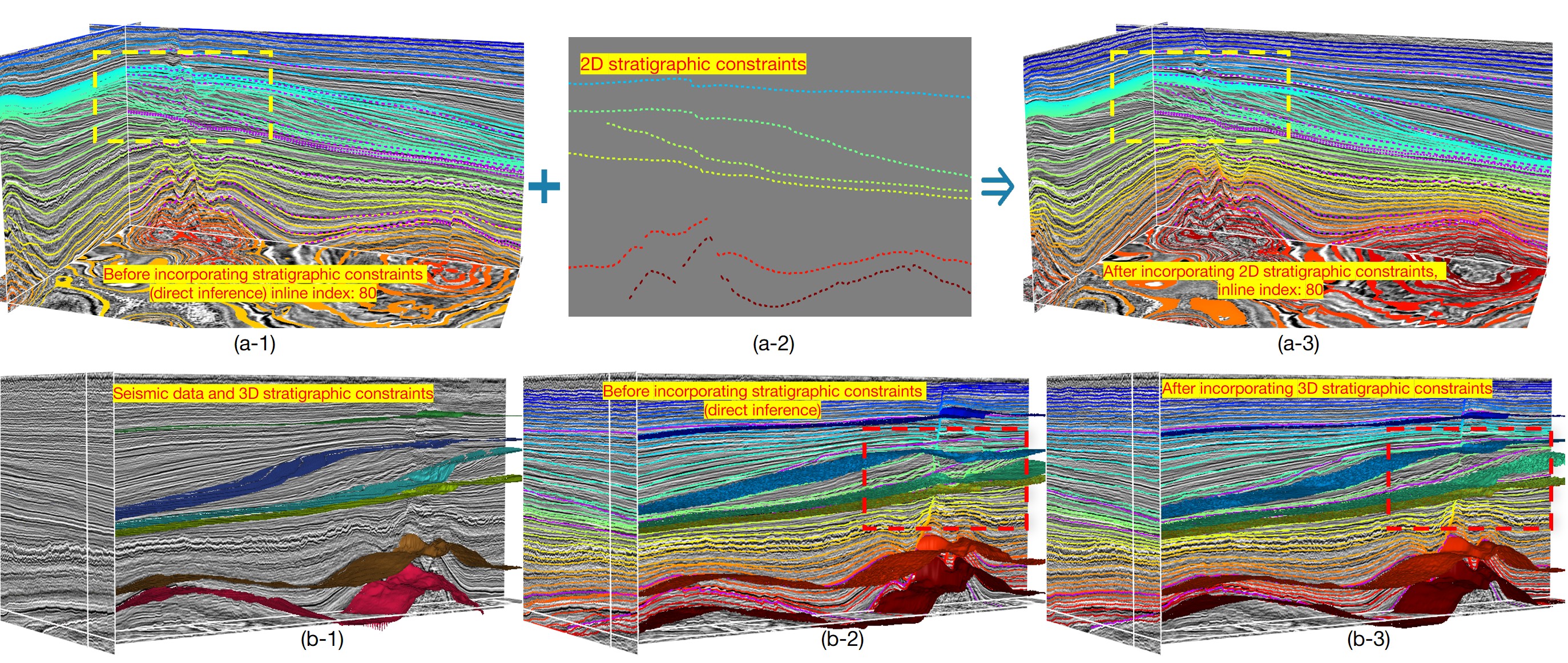}
\caption{Effect of stratigraphic constraints on RGT estimation. The purple lines represent the input 2D horizon constraints, which also serve as reference lines for the horizons. (a) Incorporating 2D horizon constraints into RGT-Est. The purple dashed curves denote ground-truth horizons used for visual comparison, and the yellow boxes highlight regions where the constrained result better honors the target stratigraphic geometry. (b) Incorporating sparse 3D horizon constraints into RGT-Est. The input seismic volume and 3D horizon constraints are shown on the left, while the direct inference result and the constraint-guided result are shown in the middle and on the right, respectively. Both 2D and 3D constraints significantly improve horizon alignment and spatial consistency of the estimated RGT field.}
\label{fig:constraint_guided}
\end{figure*}

Figure~\ref{fig:constraint_guided} evaluates the effect of incorporating sparse stratigraphic constraints into RGT-Est.
The experiment demonstrates that the proposed framework supports both direct inference and constraint-guided inference, allowing user-provided 2D or 3D horizon constraints to refine the estimated RGT field.

Figure~\ref{fig:constraint_guided}a shows the result obtained with 2D stratigraphic constraints.
In the direct inference result, RGT-Est recovers the main stratigraphic framework and produces RGT contours that generally follow the seismic reflections.
However, local deviations can still be observed in structurally complex intervals, especially where the reflectors are weak, discontinuous, or laterally variable.
The purple dashed curves denote the ground-truth horizons used for visual comparison, and the yellow boxes highlight representative regions where the direct inference result shows local mismatch.
After incorporating the 2D stratigraphic constraints, the predicted RGT contours become better aligned with the reference horizons and more consistent with the seismic reflectors.
The improvement is particularly evident in the highlighted regions, where the constrained result reduces local drift and better honors the target stratigraphic geometry.

Figure~\ref{fig:constraint_guided}b further evaluates the effect of 3D stratigraphic constraints.
The input consists of the seismic volume and sparse 3D horizon surfaces, as shown in Fig.~\ref{fig:constraint_guided}b-1.
Before incorporating these constraints, the direct inference result captures the overall stratigraphic trend, but local inconsistencies remain in regions with rapid lateral changes in reflector geometry.
In the red-box region, the estimated contours show noticeable mismatch with the stratigraphic surfaces and reduced lateral consistency.
After introducing the 3D constraints, the RGT field becomes more consistent with the input horizon surfaces and shows improved spatial continuity across the volume.
The constrained result better preserves the relative ordering between adjacent strata and reduces local distortions in the highlighted complex zone.

Overall, the experiment demonstrates that RGT-Est can flexibly support 2D or 3D horizon constraints as stratigraphic prompts.
These constraints significantly improve the accuracy of RGT estimation, the alignment between RGT contours and seismic reflectors, and the volumetric consistency of the predicted stratigraphic field.
This capability enables RGT-Est to operate not only as a fully automatic inference model but also as an interactive interpretation tool that can incorporate interpreter-provided stratigraphic priors for higher-precision RGT construction.

\subsection{Generalization on Challenging Field Surveys}
\label{sec:exp_challenging}

\begin{figure*}[!htb]
\includegraphics[width=14.3cm]{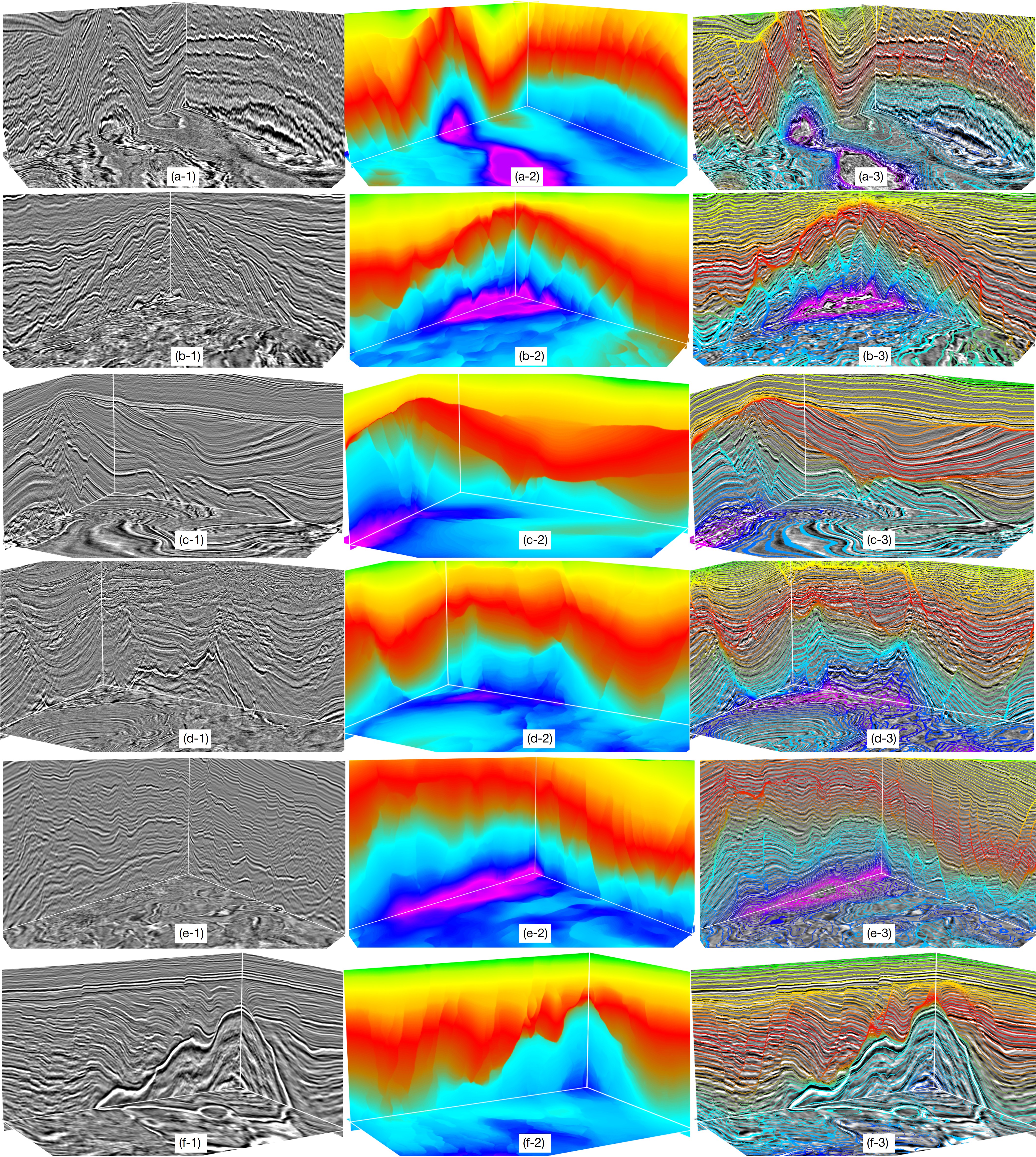}
\caption{Representative RGT estimation results on challenging field surveys. From left to right, each row shows the input seismic volume, the estimated RGT field, and the horizons extracted from the estimated RGT field and overlaid on the seismic volume. The six rows correspond to the Costa Rica survey, the Poseidon survey in Australia, two Netherlands surveys, and two field surveys from a region in China. These examples cover complex geological settings, including strong deformation, steeply dipping reflectors, faulted structures, multi-stage stratigraphic units, and diapiric or intrusive structures. RGT-Est produces coherent RGT fields and contours that generally follow the seismic reflectors across these challenging surveys.}
\label{fig:challenging_surveys}
\end{figure*}

Figure~\ref{fig:challenging_surveys} presents representative RGT estimation examples on several challenging field surveys from different geological settings.
From left to right, each row shows the input seismic volume, the seismic volume overlaid with RGT contours, and the estimated RGT field.
These examples include the Costa Rica survey, the Poseidon survey from Australia, two Netherlands surveys, and two field surveys from a region in China.
The tested volumes cover diverse structural and depositional conditions, including strong deformation, steeply dipping reflectors, complex fault systems, unconformity-like terminations, intrusive or diapiric structures, and laterally variable stratigraphic patterns.

The first row (Fig.~\ref{fig:challenging_surveys}a) shows the Costa Rica survey, where the seismic image contains strongly deformed reflectors and laterally discontinuous events.
Such structural complexity makes it difficult to construct a globally ordered RGT field, because local reflector discontinuities can easily lead to contour distortion or stratigraphic inconsistency.
RGT-Est produces a continuous RGT volume that follows the main seismic events across the deformed zones.
The overlaid contours remain generally consistent with the reflector geometry, indicating that the method can recover a coherent stratigraphic framework even when the seismic reflections are locally disrupted.

The second row (Fig.~\ref{fig:challenging_surveys}b) shows the Poseidon survey from Australia.
This example is characterized by a prominent large-scale structural high and strongly varying reflector dips.
The estimated RGT field captures the overall uplift geometry and maintains a smooth age progression from shallow to deep intervals.
In the contour-overlaid view, the predicted RGT contours conform well to the inclined reflectors on both flanks of the structure, suggesting that RGT-Est can adapt to complex three-dimensional structural relief and strong lateral variation in stratigraphic geometry.

The third row (Fig.~\ref{fig:challenging_surveys}c) shows a Netherlands survey with laterally continuous but highly variable reflectors, local deformation, stratigraphic terminations, and multi-stage depositional units.
In this area, different stratigraphic packages are interleaved and locally overprinted, producing complex reflector relationships and strong lateral variations in seismic texture.
Even under such multi-phase stratigraphic interference, RGT-Est generates a stable and coherent RGT field.
The estimated contours remain well aligned with the seismic reflections and preserve a consistent relative ordering across different stratigraphic units, indicating that the method can robustly handle complex depositional superposition and laterally varying geological architectures.

The fourth row (Fig.~\ref{fig:challenging_surveys}d) shows another Netherlands survey with dense faulting, strong lateral structural variation, and poorly defined reflection events.
In this area, the seismic reflections are weak, discontinuous, and locally obscured by fault-related deformation, making it difficult to identify continuous reflectors and maintain consistent RGT ordering across fault-bounded blocks.
Despite these ambiguous seismic responses, RGT-Est produces contours that generally follow the main reflector trends on both sides of the fault zones and preserve a coherent stratigraphic framework across the volume.
This result indicates that the proposed method can robustly handle unclear reflection axes and fault-related discontinuities while maintaining a stable relative geologic time field in the surrounding strata.

The fifth and sixth rows (Fig.~\ref{fig:challenging_surveys}e,f) show two examples from a field survey in China.
The fifth row contains a densely faulted structural setting, where steep faults cut through the stratigraphic sequence and introduce abrupt discontinuities in the seismic image.
RGT-Est successfully represents these discontinuities while maintaining a consistent ordering of the surrounding strata.
The contours follow the main reflectors on both sides of the faulted zones and avoid severe crossing or inversion artifacts.
The sixth row presents a more challenging case with a pronounced diapiric or intrusive structure, where the surrounding strata are strongly uplifted and deformed.
Despite the severe structural disturbance, RGT-Est recovers a geologically reasonable RGT field.
The contour-overlaid result shows that the predicted RGT contours bend consistently around the central structure and remain aligned with the deformed reflectors.

Overall, these examples demonstrate the strong field-data generalization ability of RGT-Est across different surveys, geological settings, and data qualities.
The method can produce coherent RGT fields not only in relatively continuous stratigraphic intervals but also in challenging regions with strong deformation, steep dips, fault-related discontinuities, multi-stage stratigraphic interference, and complex structural relief.
The consistent alignment between the estimated RGT contours and seismic reflectors indicates that RGT-Est effectively preserves stratigraphic ordering and provides a reliable basis for horizon correlation and subsequent structural interpretation.

\subsection{Feature Visualization}
\label{sec:feature_visualization}

\begin{figure*}[t]
\includegraphics[width=17cm]{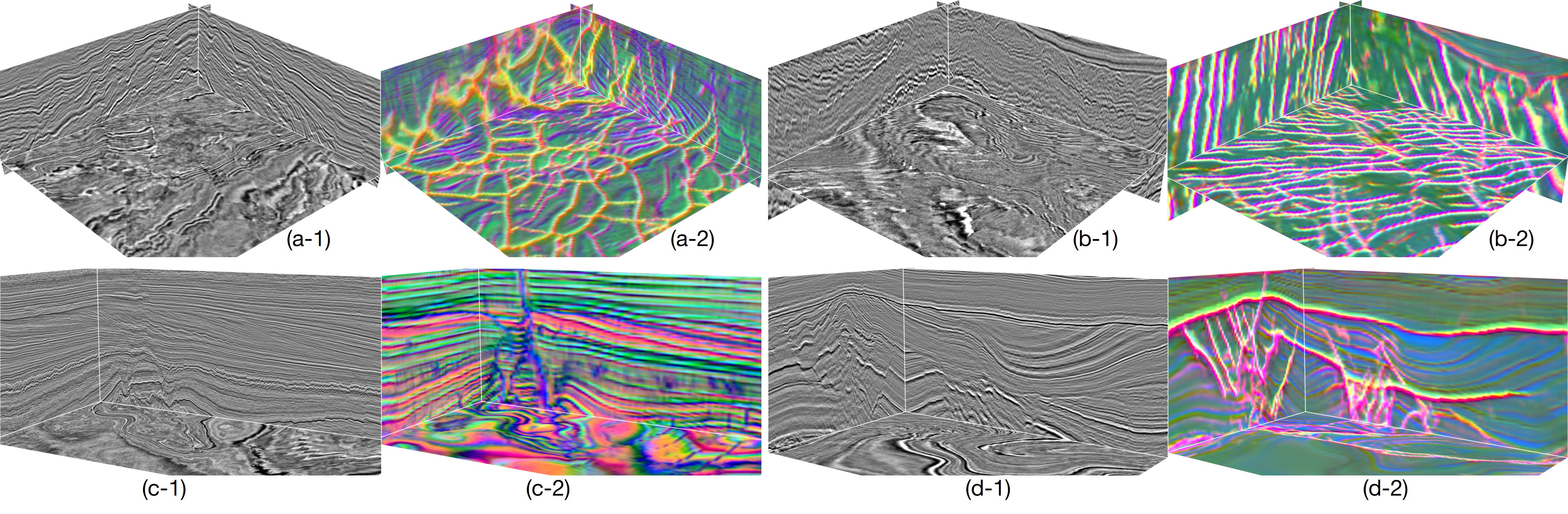}
\caption{Visualization of intermediate features learned by RGT-Est. For each example, the seismic volume is shown together with the RGB visualization of the Stage~3 HRNet feature map after PCA along the channel dimension. The feature responses highlight geologically important structures, including faults, unconformities, horizons, slope bodies, and deformed stratigraphic intervals. This indicates that RGT-Est learns structure-aware and stratigraphy-aware representations that support coherent RGT estimation.}
\label{fig:feature_pca}
\end{figure*}

To further examine what RGT-Est learns from seismic data, we visualize the intermediate feature representations extracted from Stage~3 of the HRNet backbone.
Specifically, principal component analysis is applied along the channel dimension of the feature maps, and the first three principal components are rendered as RGB volumes.
This visualization provides an intuitive way to inspect whether the learned features correspond to geologically meaningful structures rather than only low-level amplitude textures.

Figure~\ref{fig:feature_pca} shows representative examples from different field surveys.
For each case, the seismic volume is shown together with the RGB visualization of the corresponding intermediate features.
The PCA-based feature maps exhibit strong responses around key geological structures, including faults, unconformities, continuous horizons, slope bodies, and deformed stratigraphic intervals.
In faulted regions, the feature responses highlight discontinuous zones and fault-related reflector offsets.
Near unconformities and slope structures, the responses delineate major stratigraphic boundaries and lateral changes in depositional geometry.
In relatively continuous intervals, the features remain organized along reflector-parallel patterns, indicating sensitivity to stratigraphic layering.

These observations suggest that the HRNet backbone in RGT-Est learns structure-aware and stratigraphy-aware representations from seismic data.
The network does not merely encode local amplitude variations, but also captures geological elements that are critical for constructing a coherent RGT field.
Such feature organization helps explain why RGT-Est can preserve reflector alignment, maintain stratigraphic ordering, and generalize to structurally complex field surveys.

\section{Discussion}
\label{sec:discussion}

The results show that RGT-Est provides a stratigraphically consistent way to estimate relative geologic time from 3D seismic data.
Rather than treating RGT estimation as a purely numerical regression problem, the proposed method represents the RGT field in a sinusoidal phase space that is more closely related to the periodic and ordered nature of stratigraphic layering.
This design is important for solid Earth applications because an RGT volume is not only a mathematical scalar field, but also a compact geological representation of depositional order, structural deformation, and stratigraphic correlation.
When the estimated contours follow seismic reflectors and maintain a stable relative ordering, they provide a physically meaningful basis for interpreting horizons, unconformities, faults, and deformed stratigraphic packages.

The comparison with DeepRGT$^\dagger$ demonstrates the importance of this stratigraphy-aware formulation.
Voxel-space regression can recover the first-order depth trend, but it often produces local smoothing or contour distortion where seismic reflections are weak, discontinuous, dipping, or structurally disturbed.
Such errors are not merely visual artifacts.
Because RGT is used to extract horizons and correlate stratigraphic surfaces, local inconsistencies may propagate into incorrect horizon tracking, distorted depositional geometries, or unrealistic structural models.
By optimizing in the sinusoidal space and combining pointwise, perceptual, and adversarial constraints, RGT-Est produces contours that better honor seismic reflector geometry and preserve more coherent stratigraphic ordering.
This improvement is particularly relevant for structurally complex basins, where horizon continuity is commonly interrupted by faults, unconformities, slope deposits, and multi-stage deformation.

The field examples further indicate that RGT-Est can generalize across different geological settings.
The Costa Rica example contains strong deformation and laterally discontinuous reflections, the Poseidon example contains pronounced structural relief and steeply varying reflector dips, the Netherlands examples include slope structures and multi-stage stratigraphic units, and the China examples include densely faulted intervals and diapiric or intrusive structures.
These cases represent common but challenging situations in seismic interpretation, where local seismic attributes alone are often insufficient to construct a globally consistent stratigraphic framework.
The ability of RGT-Est to produce coherent RGT fields across these surveys suggests that the method learns stratigraphic organization beyond survey-specific amplitude patterns.
This property is essential for applying AI-based interpretation methods to field data from different basins, acquisition conditions, and tectono-stratigraphic environments.

The constraint-guided experiments show another practically important aspect of the method.
In seismic interpretation, fully automatic RGT estimation is useful for rapid regional screening and preliminary structural analysis, but high-precision geological interpretation often requires expert control through picked horizons or stratigraphic markers.
RGT-Est naturally supports this workflow by accepting sparse 2D or 3D horizon constraints as stratigraphic prompts.
The 2D constraints improve local alignment on the constrained section, whereas the 3D constraints provide volumetric control and improve spatial consistency across the seismic volume.
This capability links data-driven prediction with interpreter-guided geological modeling.
It also reduces the gap between automatic AI inference and traditional interpretation practice, where human expertise remains essential for resolving ambiguous seismic responses.

From a geological perspective, the estimated RGT volumes can support several downstream solid Earth applications.
First, they provide a dense stratigraphic coordinate system from which horizons can be extracted automatically and compared across faults, folds, and unconformities.
Second, they can help characterize depositional architectures such as slope systems, clinoform-like geometries, and laterally varying stratigraphic packages.
Third, they can serve as structural constraints for implicit geological modeling, reservoir characterization, and basin-scale stratigraphic analysis.
Because the RGT field encodes relative ordering rather than only local reflector appearance, it provides a useful intermediate representation between seismic images and higher-level geological models.

Several limitations should also be noted.
First, the reliability of RGT-Est still depends on the quality of seismic imaging.
In regions where reflectors are severely obscured by noise, multiples, acquisition footprint, poor illumination, or strong velocity-related distortion, the predicted RGT field may remain uncertain.
Second, although the sinusoidal representation improves stratigraphic consistency, the present framework does not explicitly impose hard geological constraints, such as strict monotonicity, non-intersection of isosurfaces, or fault-specific displacement relationships.
Third, sparse horizon constraints can improve the result, but their effectiveness depends on the accuracy, distribution, and geological representativeness of the provided constraints.
Future work could incorporate uncertainty estimation, fault-aware structural constraints, well markers, and physics-informed geological priors to further improve the reliability of RGT estimation in highly complex tectonic settings.

Overall, RGT-Est provides a practical framework for learning stratigraphically consistent RGT volumes from 3D seismic data.
By combining a stratigraphy-aware sinusoidal representation with optional horizon guidance, the method supports both automatic and interpreter-guided workflows.
This makes it useful not only for improving AI-based RGT estimation, but also for advancing seismic interpretation toward more consistent, reproducible, and geologically meaningful structural and stratigraphic modeling.

\conclusions
\label{sec:summary}

We developed RGT-Est, a deep learning framework for estimating relative geologic time from 3D seismic data.
The method maps the predicted RGT field into a multi-frequency sinusoidal phase space and jointly uses pointwise, perceptual, and adversarial losses to improve stratigraphic consistency and geological plausibility.
Experiments on multiple field surveys show that RGT-Est produces RGT contours that better follow seismic reflectors and preserve more coherent stratigraphic ordering than voxel-space regression.
The method remains effective in challenging settings, including weak-reflection zones, slope structures, faulted intervals, multi-stage stratigraphic units, and diapiric or intrusive structures.
RGT-Est also supports sparse 2D or 3D horizon constraints as stratigraphic prompts, which significantly improve horizon alignment and volumetric consistency.
These results indicate that sinusoidal-space optimization provides an effective and flexible strategy for constructing geologically meaningful RGT volumes for horizon extraction, structural interpretation, and subsequent geological modeling.

%% The following commands are for the statements about the availability of data sets and/or software code corresponding to the manuscript.
%% It is strongly recommended to make use of these sections in case data sets and/or software code have been part of your research the article is based on.

\codedataavailability{
	The source code of RGT-Est is openly available at \url{https://github.com/douyimin/RGT-Est}, 
	and the associated trained model weights and seismic datasets used to reproduce the results in this work are archived on Zenodo at \url{https://doi.org/10.5281/zenodo.20118902} \citep{rgtest2026}.
}

\authorcontribution{Y.D. and X.W. conceived the study and designed the methodology. Y.D. implemented the framework, conducted the experiments, and prepared the figures. X.W. supervised the project and provided guidance on the seismic interpretation aspects. H.G. and Z.B. contributed to the discussion of methodology, data analysis, and result validation. All authors contributed to writing and revising the manuscript.}

\competinginterests{The authors declare that they have no conflict of interest.}

%\begin{acknowledgements}
%The authors thank the providers of the open-access seismic datasets used in this study, including the Costa Rica fore-arc basin survey, the Poseidon survey from Australia, and the Netherlands surveys (e.g., F3). The authors also thank the developers of open-source tools that made this work possible.
%\end{acknowledgements}

%% REFERENCES

\bibliographystyle{copernicus}
\bibliography{refs}

\end{document}